\documentclass[hidelinks,journal,onecolumn]{IEEEtran}
\IEEEoverridecommandlockouts

\usepackage{cite}
\usepackage{amsmath,amssymb,amsfonts}
\usepackage{graphicx}
\usepackage{textcomp}
\usepackage{xcolor}
\usepackage[acronym,shortcuts]{glossaries}
\usepackage{bm}
\usepackage{caption}
\usepackage{subcaption}
\usepackage{relsize}
\usepackage{soul}
\usepackage{algorithm, algpseudocode}
\usepackage{outlines}
\usepackage{isomath}
\usepackage{lipsum}
\usepackage{orcidlink}
\usepackage{dblfloatfix}
\usepackage[bottom]{footmisc}
\usepackage{comment}
\usepackage{mleftright}

\def\BibTeX{{\rm B\kern-.05em{\sc i\kern-.025em b}\kern-.08em
T\kern-.1667em\lower.7ex\hbox{E}\kern-.125emX}}

\newcommand{\trans}[0]{^{\mathsf{T}}}
\newcommand{\transs}[0]{^{\!\mathsf{T}}}

\newacronym{OS-QSM}{OS-QSM}{optimised scalable QSM}
\newacronym{GSM}{GSM}{generalised spatial modulation}
\newacronym{FDFR}{FDFR}{full-diversity full-rate}
\newacronym{mMIMO}{mMIMO}{massive multiple-input multiple-output}
\newacronym{MIMO}{MIMO}{multiple-input multiple-output}
\newacronym{MU}{MU}{multi-user}
\newacronym{OFDM}{OFDM}{orthogonal frequency-domain multiplexing}
\newacronym{IM}{IM}{index modulation}
\newacronym{IoT}{IoT}{Internet-of-Things}
\newacronym{QSM}{QSM}{quadrature spatial modulation}
\newacronym{BP}{BP}{belief propagation}
\newacronym{GaBP}{GaBP}{Gaussian belief propagation}
\newacronym{SM}{SM}{spatial modulation}
\newacronym{IQ}{IQ}{in-phase and quadrature}
\newacronym{ML}{ML}{machine learning}
\newacronym{BER}{BER}{bit error rate}
\newacronym{P2P}{P2P}{point-to-point}
\newacronym{AWGN}{AWGN}{additive white Gaussian noise}
\newacronym{SWIPT}{SWIPT}{simultaneous wireless information and power transfer}
\newacronym{CSI}{CSI}{channel state information}
\newacronym{MQAM}{$M$-QAM}{$M$-ary quadrature amplitude modulation}
\newacronym{IC}{IC}{interference cancellation}
\newacronym{SGA}{SGA}{scalar Gaussian approximation}
\newacronym{CLT}{CLT}{central limit theorem}
\newacronym{PDF}{PDF}{probability density function}
\newacronym{GB-ISTA}{GB-ISTA}{greedy boxed iterative soft-thresholding algorithm}
\newacronym{MP}{MP}{message passing}
\newacronym{WL}{WL}{wireless localization}
\newacronym{SD}{SD}{sphere decoder}
\newacronym{FD}{FD}{full-duplex}
\newacronym{STC}{STC}{space-time coding}
\newacronym{SotA}{SotA}{state-of-the-art}
\newacronym{IER}{IER}{index vector error rate}
\newacronym{B5G}{B5G}{beyond fifth generation}
\newacronym{STC-SM}{STC-SM}{space-time coded SM}
\newacronym{STC-QSM}{STC-QSM}{space-time coded QSM}
\newacronym{SC-IM}{SC-IM}{single-carrier IM}
\newacronym{SSK}{SSK}{space shift keying}
\newacronym{mmWave}{mmWave}{millimeter-wave}
\newacronym{THz}{THz}{Terahertz}
\newacronym{RIS}{RIS}{reflective intelligence surface}
\newacronym{RF}{RF}{radio frequency}
\newacronym{STBC}{STBC}{space-time block code}
\newacronym{MMSE}{MMSE}{minimum mean-squared-error}
\newacronym{CS}{CS}{compressive sensing}
\newacronym{i.i.d.}{i.i.d.}{independent and identically distributed}
\newacronym{JCAS}{JCAS}{joint communication and sensing}
\newacronym{ISAC}{ISAC}{integrated sensing and communication}
\newacronym{JRC}{JRC}{joint radar-communications}
\newacronym{SE}{SE}{spectral efficiency}
\newacronym{EE}{EE}{energy efficiency}
\newacronym{RBL}{RBL}{rigid body localization}
\newacronym{RBT}{RBT}{rigid body tracking}
\newacronym{SC-RBL}{SC-RBL}{soft-connected RBL}
\newacronym{W-RBL}{W-RBL}{\underline{wireless} RBL}
\newacronym{GA}{GA}{genie-aided}
\newacronym{MC}{MC}{matrix completion}
\newacronym{EA}{EA}{``\emph{estimate-then-average}''}
\newacronym{AE}{AE}{``\emph{average-then-estimate}''}
\newacronym{IRS}{IRS}{intelligent reflecting surface}
\newacronym{RSSI}{RSSI}{received signal strength indicator}
\newacronym{PSO}{PSO}{particle swarm optimization}

\newacronym{NTN}{NTN}{non-terrestrial networks} 
\newacronym{6G}{6G}{sixth-generation}
\newacronym{3D}{3D}{three-dimensional}
\newacronym{D2D}{D2D}{device-to-device}
\newacronym{RR}{RR}{round-robin}
\newacronym{DA}{DA}{Dutch auction}
\newacronym{CWFL}{CWFL}{clustered WFL}
\newacronym{WFL}{WFL}{wireless federated learning}
\newacronym{RSMA}{RSMA}{rate splitting multiple access}

\newacronym{TDMA}{TDMA}{time-domain multiple access}
\newacronym{NOMA}{NOMA}{non-orthogonal multiple access}

\newacronym{CT}{CT}{compute-then-transmit}
\newacronym{SDP}{SDP}{semidefinite programming}
\newacronym{FP}{FP}{fractional programming}
\newacronym{CF-mMIMO}{CF-mMIMO}{cell free massive MIMO}
\newacronym{iid}{i.i.d.}{independent and identically distributed}
\newacronym{DL}{DL}{downlink}
\newacronym{UL}{UL}{uplink}
\newacronym{MDS}{MDS}{multidimensional scaling}

\newacronym{SIC}{SIC}{successive interference cancellation}
\newacronym{BS}{BS}{base station}
\newacronym{TX}{TX}{transmit}
\newacronym{RX}{RX}{receive}

\newacronym{SISO}{SISO}{single-input single-output}

\newacronym{SINR}{SINR}{signal-to-interference-and-noise ratio}
\newacronym{FL}{FL}{federated learning}
\newacronym{CPU}{CPU}{central processing unit}
\newacronym{KNN}{KNN}{K-nearest-neighbor}

\newacronym{GD}{GD}{gradient descent}

\newacronym{RSS}{RSS}{received signal strength}
\newacronym{FIM}{FIM}{fisher information matrix}
\newacronym{ToA}{ToA}{time of arrival}
\newacronym{AoA}{AoA}{angle of arrival}
\newacronym{GP}{GP}{Gaussian process}
\newacronym{2D}{2D}{two-dimensional}
\newacronym{GPR}{GPR}{Gaussian process regression}
\newacronym{GNSS}{GNSS}{global navigation satellite systems}

\newacronym{RRH}{RRH}{remote radio head}
\newacronym{GPS}{GPS}{Global Positioning System}
\newacronym{RFID}{RFID}{radio frequency identification}
\newacronym{TCAS}{TCAS}{traffic alert and collision avoidance systems}
\newacronym{RMSE}{RMSE}{root mean square error}
\newacronym{SGD}{SGD}{stochastic gradient descent}

\newacronym{CU}{CU}{computing unit}
\newacronym{DM-MIMO}{DM-MIMO}{distributed massive multiple-input multiple-output}
\newacronym{LOS}{LOS}{line-of-sight}
\newacronym{NLOS}{NLOS}{non-line-of-sight}
\newacronym{ROI}{ROI}{region of interest}
\newacronym{AP}{AP}{access point}
\newacronym{PoCs}{PoCs}{projections onto convex sets}
\newacronym{TDOA}{TDOA}{time difference of arrival}
\newacronym{DoA}{DoA}{direction of arrival}
\newacronym{UE}{UE}{user equipment}
\newacronym{dB}{dB}{decibel}

\newacronym{CG}{CG}{conjugate gradient}
\newacronym{SC}{SC}{soft-connected}
\newacronym{CRLB}{CRLB}{Cramér-Rao Lower Bound}
\newacronym{PoA}{PoA}{phase of arrival}
\newacronym{UAV}{UAV}{unmanned aerial vehicle}
\newacronym{VR}{VR}{virtual reality}

\begin{document}

\title{Soft-connected Rigid Body Localization:\\ State-of-the-Art and Research Directions for 6G \\[-0.75ex]

{\normalsize \textit{White paper for SPM Special Issue on "Signal Processing for the Integrated Sensing and Communications Revolution".}} \\}

\author{\normalsize 
Niclas Führling$^{1}$,
Hyeon Seok Rou$^{1}$,
Giuseppe Thadeu Freitas de Abreu$^{1}$,
David Gonz{\'a}lez G.$^{2}$, and 
Osvaldo Gonsa$^{2}$.
\\[2ex]

{
\small $^1$School of Computer Science and Engineering, Constructor University, 28759 Bremen, Germany \\[-0.5ex]

$^{2}$Wireless Communications Technologies, Continental AG, 65936 Frankfurt am Main, Germany
}}

\maketitle

\vspace{-7ex}
\section{\textbf{Author Biography}}
\par 

\textbf{Niclas Führling} \textit{(Graduate Student Member, IEEE)} [nfuehrling@constructor.university] received the B.Sc. degree in electrical and computer engineering from Jacobs University Bremen, Bremen, Germany in 2022. He is currently pursuing the M.Sc. degree in electrical engineering with the University of Bremen, with a focus on communication and information technology, while working on a research project at Constructor University, focusing on 6G connectivity. His current research interests are wireless communications and signal processing.

\textbf{Hyeon Seok Rou} \textit{(Graduate Student Member, IEEE)} [hrou@constructor.university]  is a Ph.D. Candidate at Constructor University, Bremen, Germany, funded as a Research Associate at Continental AG on a researach project on 6G vehicular-to-everything (V2X) integrated sensing and communications (ISAC).
His research interests lie in the fields of ISAC, hyper-dimensional sparse modulation schemes, B5G/6G V2X wireless communications technology, and Bayesian inference.

\textbf{Giuseppe Thadeu Freitas de Abreu} \textit{(Senior Member, IEEE)} [gabreu@constructor.university] is a Full Professor of Electrical Engineering at Constructor University, Bremen, Germany. His research interests include communications theory, estimation theory, statistical modeling, wireless localization, cognitive radio, wireless security, MIMO systems, ultrawideband and millimeter wave communications, full-duplex and cognitive radio, compressive sensing, energy harvesting networks, random networks, connected vehicles networks, and many other topics. He has served as an editor for various IEEE Transactions, and currently serves as an editor to the IEEE Signal Processing Letters and the IEEE Communications Letters. 

\textbf{David González G.} \textit{(Senior Member, IEEE)} [david.gonzalez.g@ieee.org] %
is a Senior Research Engineer at Continental AG, Germany, and has previously served at Panasonic Research and Development Center, Germany.
His research interests include aspects of cellular networks and wireless communications, including interference management, radio access modeling and optimization, resource allocation, and vehicular communications. 
Since 2017, he has represented his last two companies as delegate in the 3GPP for 5G standardization, mainly focused on physical layer aspects and vehicular communications.

\textbf{Osvaldo Gonsa} [osvaldo.gonsa@continental-corporation.com] received the Ph.D. degree in electrical and computer engineering from Yokohama National University, Japan, in 1999, and the M.B.A. degree from the Kempten School of Business, Germany, in 2012. Since 1999 he has worked in research and standardization in the areas of core and radio access network. He is currently the Head of the Wireless Communications Technologies Group, Continental AG, Frankfurt, Germany. And since 2020 also serves as a member for the GSMA Advisory Board for automotive and the 6GKom Project of the German Federal Ministry of Education and Research.

%

\section{\textbf{History, motivation, and significance of the topic}}

\Ac{WL} (aka positioning) is an early example of \ac{ISAC} that demonstrates how communication signals can also be used to extract environmental information ($i.e.$, the location of users) using signals originally indented for communications \cite{LimaAccess2021, KwonJSTSP2021, johannsen2022joint}, and which has proven highly successful in a wide range of applications including security \cite{ZhengTDSC2017}, smart homes \cite{BianchiTIM2019}, industrial automation \cite{BarbieriTIM2021}, robotics \cite{WangRAM2021}, vehicular networks \cite{ChuTVT2021}, and more.

In terms of fundamental concept, $i.e.$, the type of information extracted from the wireless links for the purpose of localization, many \ac{SotA} techniques have been proposed over the years, including (to cite only a few), methods based on finger-prints \cite{VoCST2016}, \ac{RSSI} \cite{LamTVT2019}, \ac{AoA} \cite{Al-SadoonTAP2020}, radio range \cite{ZengTSP2022}, and even mere connectivity \cite{ShangTPDS2004} or hop-counts \cite{RahmatollahiTC2012}, as well as hybrid combinations of the latter \cite{GhodsTWC2018,MacagnanoTWC2013}.

The \ac{WL} literature also exhibits vast diversity in terms of the underlying mathematical approaches employed in the design of positioning algorithms, with examples ranging from purely algebraic methods exploiting \ac{MDS} \cite{GhodsTWC2018,MacagnanoTWC2013, NorooziWCL2018}, to methods based on optimization-theoretical techniques such as \ac{SDP} \cite{BiswasTASE2006}, \ac{PoCs} \cite{ZhangTWC2015}, \ac{FP} \cite{WangAccess2018} and \ac{MC} \cite{NguyenTC2019, IimoriAsilomarMC2020, LiuTAES2023}, to schemes based on \ac{ML} \cite{Prasad_2018, NessaAccess2020, SinghAccess2021}.

Finally, in addition to the aforementioned variety of concepts and mathematical approaches, abundance also exists in the \ac{WL} literature in terms of offered features, $i.e.$, the various challenges addressed, which besides issues such as robustness against noise, bias, \ac{LOS} and \ac{NLOS} conditions already covered by the articles cited above, include also the mitigation of effects such as information scarcity \cite{DestinoTWC2009}, uncertainty of anchor points \cite{VeldeJSAC2015}, co-existence of near- and far-field waves \cite{WangTWC2019} and more.

Despite the healthy breadth of topics covered by the literature, one aspect of the problem that has been largely ignored by most \ac{SotA} \ac{WL} methods is the fact that, in many applications, targets are generally not single points, but rather \ac{3D} objects.
Aiming to address this matter, a growing literature is emerging, in which each target is modeled not as a single point, but as a group of inter-connected points with a fixed arrangement, $i.e$, a rigid body \cite{ChepuriTSP2014, Zhou_2019, WangTSP2020}.

Thanks to the inherent inclusion of target shape and orientation models, which can be either used as prior information or jointly estimated within the solution of the localization problem, these \ac{RBL} techniques, have been shown to exhibit higher accuracy than ``standard'' (target point-based) \ac{WL} methods.
In view of the latter, an increasing number of contributions aimed at building onto standard \ac{WL} schemes towards \ac{RBL} variations has appeared, which often utilize the same concepts and approaches of the latter, and which aim to exhibit similar features.
Examples (to cite only a few) include the \ac{MDS}-based multi-target \ac{RBL} scheme of \cite{PizzoICASSP2016}; the range-based and information-heterogeneous \ac{SDP}-based approaches of \cite{JiangTWC2019} and \cite{Wu_22}, respectively; schemes that seek to equip \ac{RBL} with robustness to \ac{NLOS} conditions \cite{Wan_21}, anchor uncertainty \cite{ZhaISEEIE2021} and errors from imperfect synchronization \cite{DongTWC2023}; and the \ac{MC}- and \ac{ML}-based \cite{AnTVT2023, YangJEET2020} approaches to address missing information.

Emphasizing the distinction between \ac{WL}/\ac{ISAC} \cite{LimaAccess2021}, where wireless 
connectivity is a fundamental component of the technology, and other approaches based on $e.g.$ inertial sensing \cite{Chen2023deep} or machine vision \cite{XinRICAI2019}, which are outside of the scope of this article, and with pardon for the oversimplification, it can be said based on the above that with respect to \ac{RBL}, the current focus of the \ac{WL}/\ac{ISAC} research community is essentially extending the multitude of techniques found in the vast \ac{ML} literature to the \ac{RBL} problem,  which could perhaps be summarized in the term \ac{W-RBL}.

In view of the above, {\bf the first intended contribution of the full-version of this article is  a compact but comprehensive survey of \ac{WL} algorithms from the perspective of their extension to the \ac{W-RBL} paradigm}, it in terms of their concepts, approaches and features, and with emphasis on their role to \ac{ISAC} and \ac{6G} systems \cite{Trevlakis2023localization}.
In addition, we also consider the perspective that in key applications of \ac{W-RBL} as a \ac{6G}/\ac{ISAC} technology, such as  autonomous vehicles \cite{ChalvatzarasTIV2023}, industrial automation \cite{Alwis6G2023}, and \ac{NTN} \cite{DureppagariArxiv2023}, objects localized or tracked are often coordinated, and/or inserted in an environmental context that implicates on constraints onto their relative location and orientation.

Examples of the former are platooning \cite{ZengTC2019} and swarming \cite{IbukiRAL2020} applications, whereby multiple vehicles/drones communicate and control their velocities and locations relative to one another, so as to maintain a given formation.
Another example is  \ac{RBL} target tracking \cite{EckenhoffRAL2019}, aka 
\ac{RBT}, whereby known or estimated velocities and orientation of targets at previous times, as well as information on physical constraints onto their trajectory ($e.g.$ the roads) inform the sequence of \ac{RBL} estimates \cite{YuITJ2023}.
These cases imply that if conformation constraints similar to those imposed onto the points of a single rigid-body are enforced for the group or sequence of targets, the entire formation can be localized simultaneously, with advantages similar to those obtained by \ac{RBL} over standard \ac{WL} approaches.

{\bf We refer to this perspective as \ac{SC-RBL}, and a discussion of mechanisms to extend \ac{RBL} techniques to \ac{SC-RBL} is another intended contribution of the full version of the article.}

\section{\textbf{Outline of the proposed Special Issue paper}}

%
\label{sec:SotA}
%


In light of the above-described motivation to evolve from point-based localization methods to \ac{RBL} methods, this section will aim to provide the mathematical analysis and derivation of formulating the \ac{RBL} problem from the classical point-based localization methods.
Specifically, to illustrate, we highlight a variety of \ac{SotA} point-based localization methods, and develop the mathematical formulation into the \ac{RBL} framework.
The full article is expected include a more elaborate mathematical analysis and a thorough survey and development of \ac{SotA} methods.

\vspace{-1ex}
\subsection{\textbf{Background of Rigid Body Localization}}
\label{sec:rb}

To that extent, the mathematics of the underlying \ac{RBL} system model is briefly introduced.
Given a rigid body, the position of the $k$-th node within a rigid body can be described by a \ac{2D} coordinate vector $\mathbf{c}_{k} \in \mathbb{R}^{2 \times 1}$,
\begin{equation}
\mathbf{c}_{k} \triangleq
\begin{bmatrix}
x_{k}, ~ y_{k}
\end{bmatrix}\trans \in \mathbb{R}^{2 \times 1},
\label{eq:coordinate_vector}
\end{equation}
where $x_{k}$ and $y_{k}$ are respectively the $x$- and $y$-coordinates of the $k$-th node of the rigid body, for all $K$ points defining the rigid body.
Next, following the \ac{SotA} system model of the standard \ac{RBL} framework \cite{PizzoICASSP2016, JiangTWC2019, Wu_22, Wan_21, ZhaISEEIE2021, DongTWC2023}, a change in position and orientation of a \ac{RBL} can be described by the joint affine transform of the $K$ coordinates of a rigid body, given by \vspace{-1ex}
\begin{equation}
\mathbf{S} \triangleq [\mathbf{s}_1, \cdots\!, \mathbf{s}_k, \cdots\!, \mathbf{s}_K] = \mathbf{R} \!\cdot\! \overbrace{[\mathbf{c}_1, \cdots\!, \mathbf{c}_k, \cdots\!, \mathbf{c}_K]}^{\triangleq \; \mathbf{C}} +\;\! [\mathbf{t}, \cdots\!, \mathbf{t}, \cdots\!, \mathbf{t}] = \mathbf{R}\mathbf{C}+ (\mathbf{t}\otimes\mathbf{1}_{1\times K}) \in \mathbb{R}^{2 \times K}, 
\label{eq:sys_mod}
\end{equation}
where $\mathbf{S} \in \mathbb{R}^{2 \times K}$ consists of the new coordinates of the rigid body nodes after the transform, $\mathbf{R}\in \mathbb{R}^{2 \times K}$ and $\mathbf{t} \in \mathbb{R}^{2 \times 1}$ respectively describe the rotation and translation equally to the $K$ original coordinates, and $\otimes$ is the kronecker product.

In addition, another important property of rigid bodies is the fixed relative Euclidean distances between the $K$ nodes in $\mathbf{C}$, which must also be satisfied in $\mathbf{S}$.
Therefore, such information is incorporated in \ac{RBL} method as a necessary constraint.

The implication for \ac{RBL} is that, the known structure of the rigid body in $\mathbf{C}$ will be exploited to aid the estimation of the localized positions in $\mathbf{S}$, where the affine transform $\mathbf{R}$ and $\mathbf{t}$ is unknown, and only some noisey observation of $\mathbf{S}$ is available.

Note that the above model and properties can easily be extended to the \ac{3D} scenario by adding another dimension to the coordinate vector in eq. \eqref{eq:coordinate_vector} and correspondingly in eq. \eqref{eq:sys_mod}, with out a loss of generality.

\subsection{\textbf{From Wireless Point-based Localization to Wireless Rigid Body Localization (W-RBL)}}
In general, there are many different approaches for wireless localization, such as \ac{MDS}, \ac{SDP}, \ac{MC}, or \ac{ML}, leveraging many different types of information, such as \ac{RSSI}, \ac{AoA}, \ac{DoA} or range.
In this next section, we first aim to describe the process of extending a standard point-based localization method to incorporate the rigid body framework in Sec. \ref{sec:rb}.
Specifically, we illustrate a derivation for the exemplified \ac{DoA}-based localization method below, followed by a brief outline of other extended \ac{RBL} approaches.
As will be revealed in the following, any localization method can effectively be reformulated to incorporate the \ac{RBL} constraints by an appropriate insertion of the \ac{RBL} framework given by eq. \eqref{eq:sys_mod}, opening significant directions of improving the existing \ac{SotA} methods.

\vspace{1ex}
\subsubsection{\textbf{Derivation of \acf{DoA}-based wireless \acf{RBL}}} $~$ 

Let us consider a \ac{2D} wireless localization scenario of a single point transmitter positioned at $\mathbf{c}^{\mathsf{TX}}$, where only the \ac{DoA} $\theta$ from the horizontal axis can be estimated at the detecting base station positioned at $\mathbf{c}^{\mathsf{BS}}$.
Then, the true \ac{DoA} $\theta$ is obtained by 
\begin{equation}
\theta = \mathrm{arccos}\big(||\mathbf{c}^{\mathsf{TX}} - \mathbf{c}^{\mathsf{BS}}||\big).
\label{eq:doa}
\end{equation}

Consequently, given a noisy observation of the \ac{DoA} denoted by $\hat{\theta}$ the \textit{observation matching} localization method aims to find the estimated coordinates $\tilde{\mathbf{c}}^{\mathsf{TX}}$ by minimizing
\begin{equation}
\tilde{\mathbf{c}}^{\mathsf{TX}} = \underset{\mathbf{c}^{\mathsf{TX}}}{\mathrm{argmin}} \; | \theta - \tilde{\theta} |^2 = \underset{\mathbf{c}^{\mathsf{TX}}}{\mathrm{argmin}} \; \big| \theta - \mathrm{arccos}(||\mathbf{c}^{\mathsf{TX}} - \mathbf{c}^{\mathsf{BS}}||) \big|^2,
\end{equation}
where the coordinates of the base station is known.

As can be seen, such single target to single base station scenario results in no diversity and no constraint, resulting in an unbounded feasible solution space.
Typically in such scenarios, multiple base stations can be leveraged to increase the estimation performance by jointly estimating the target coordinates by minimizing the mean squared error.
However, simply by considering the target of the above scenario to be a rigid body of $K$ nodes, \ac{RBL} can be leveraged to achieve a more elaborate optimization problem in which contraints is gained to provide a tighter solution space.

Specifically, consider the \ac{DoA} of eq. \eqref{eq:doa} computed for all $K$ unique nodes to yield $\theta_1,\cdots,\theta_K$.
Consequently, the minimization problem can be rewritten as \vspace{-1ex}
\begin{equation}
\tilde{\mathbf{C}}^{\mathsf{TX}} \triangleq [\tilde{\mathbf{c}}_1^{\mathsf{TX}},\cdots\!,\tilde{\mathbf{c}}^{\mathsf{TX}}_K] 
= \underset{\mathbf{c}_1^{\mathsf{TX}},\cdots,\mathbf{c}_K^{\mathsf{TX}}}{\mathrm{argmin}} \; \sum_{k=1}^{K} \Big| \theta - \mathrm{arccos}\big(||\mathbf{c}^{\mathsf{TX}}_k - \mathbf{c}^{\mathsf{BS}}||\big) \Big|^2, 
\end{equation}
subject to the rigid body constraint of fixed relative Euclidean distances between all $\mathbf{c}_1^{\mathrm{TX}},\cdots,\mathbf{c}_K^{\mathrm{TX}}$.

Since the above constraint of relative Euclidean distances is mathematically described by the affine transformation in eq. \eqref{eq:sys_mod}, the minimization problem can be rewritten as the direct estimation problem of the affine transform parameters, \textit{i.e.,} the rotation matrix of a given angle $\mathbf{R}(\alpha)$ and the translation vector $\mathbf{t} = [t_x, t_y]\trans$, \vspace{-1ex}
\begin{align}
\tilde{\mathbf{C}}^{\mathsf{TX}} 
= \underset{\mathbf{s}_1,\cdots,\mathbf{s}_K}{\mathrm{argmin}} \; \sum_{k=1}^{K} \Big| \theta - \mathrm{arccos}\big(||[\mathbf{S}]_{:,k} - \mathbf{c}^{\mathsf{BS}}||\big) \Big|^2 
= \underset{\alpha, t_x, t_y}{\mathrm{argmin}} \; \sum_{k=1}^{K} \Big| \theta - \mathrm{arccos}\big(||\overbrace{[\mathbf{R}(\alpha)\mathbf{X} + \mathbf{t}]_{:,k}}^{\triangleq \; \mathbf{s}_k} -\; \mathbf{c}^{\mathsf{BS}}||\big) \Big|^2,
\end{align}
where $[\mathbf{S}]_{:,k} \triangleq \mathbf{s}_k$ denotes the $k$-th column of the matrix $\mathbf{S}$, obtained via $\mathbf{R}(\alpha)\mathbf{X} + \mathbf{t}$, and $\mathbf{X}$ is the \textit{effective} original position of the rigid body nodes, obtained via the eigenvector decomposition of the relative distances \cite{Zhou_2019}.

In light of the above, it can be seen that the estimation problem has been transformed such that only the affine transform parameters $\alpha$ and $\mathbf{t}$ are estimated to minimized the object function, which inherently and efficiently obeys the rigid body structure constraint.
As before, the above method can also be extended to a \ac{3D} scenario, which would simply extend the coordinate vectors into $\mathbb{R}^3$, and introduce a second axis on the \ac{3D} rotation matrix $\mathbf{R}_3(\alpha, \beta)$, and $\mathbf{t} = [t_x, t_y, t_z]
\trans$.

\vspace{2ex}

Next, we provide some \ac{SotA} methods and direction of approach to enable \ac{W-RBL} in other localization scenarios with different parameters and approaches.
The full derivations and implications of each method will be addressed in the full article.

\vspace{1ex}
\subsubsection{Range-based localization} $~$ 

Multiple papers, such as \cite{Wu_22, PizzoICASSP2016} cover \ac{RBL} by leveraging range measurements, where the same system model as in \eqref{eq:sys_mod} is employed, consisting of a rotation matrix and a translation.
By using this information in a cooperative \acf{SDP} solution, the rotation and translation for the rigid bodies can be found, approaching \ac{CRLB} for low noise levels.
This is done by solving the constrained weighted least square problem formulated as
\begin{equation}
\tilde{\mathbf{y}} = \underset{\mathbf{y}}{\mathrm{argmin}}\big(\mathbf{G}\mathbf{y}-\mathbf{h}\big)\transs\mathbf{W}\big(\mathbf{G}\mathbf{y}-\mathbf{h}\big),
\end{equation}
where $\mathbf{y}$ is defined as the unknown vector including all information of the affine transform and the original positions, \textit{i.e.,} rotation angles, translation values, and original coordinate parameters, $\mathbf{G}$ and $\mathbf{h}$ are defined by certain operations of anchor positions and knowledge of the reference frame, further explained in the appendix of \cite{Wu_22}, and $\mathbf{W}$ is a weighting matrix.
In this case, the entire information of the \ac{RBL} model, including the target node positions, is concatenated into a single vector $\mathbf{y}$, which is consequently directly optimized.

\vspace{1ex}
\subsubsection{\Acf{RSSI}-based \acf{GPR} localization} $~$ 

In the following, we hint the direction of incorporating the \ac{RBL} method to a \ac{RSSI}-based \ac{GPR} localization \cite{Prasad_2018}, where $K$ targets are localized by $M$ distributed sensors via \ac{RSSI}.
Omitting the detailed derivation, the \ac{RSSI} vector collecting the \ac{RSSI} values of the $k$-th target received by the $M$ sensors, is described by $\mathbf{p}_k \in \mathbb{R}^{M} \times 1$, which is a given function of the target and sensor coordinates, and the environmental parameters such as channel fading gain.

Ultimately, the \ac{GPR} method amounts to obtaining the unknown covariance functions $\phi(\mathbf{p}_{k_1},\mathbf{p}_{k_2})$ for a pair of different \ac{RSSI} vectors $\mathbf{p}_{k_1}$ and $\mathbf{p}_{k_2}$ defined by
\begin{equation}
\phi(\mathbf{p}_{k_1},\mathbf{p}_{k_2}) = \alpha e^{-0.5(\mathbf{p}_{k_1} -\mathbf{p}_{k_2})\trans \mathbf{B}^{-1} (\mathbf{p}_{k_1}-\mathbf{p}_{k_2})} + \gamma \mathbf{p}_{k_1}\trans\mathbf{p}_{k_2} + \sigma^2_\mathrm{er}\delta_{k_1k_2},
\label{eq:covar_function}
\end{equation}
where latter $\sigma^2_\mathrm{er}\delta_{k_1k_2}$ is an error term, and the hyper parameters $\alpha, \mathbf{B} \triangleq \mathrm{diag}(\beta_1,\cdots,\beta_M), \gamma$ are to be optimized via a training phase using the proposed \ac{GPR} method.

To incorporate the \ac{RBL} framework, it must be assumed that the $K$ targets follow the same \ac{RBL} constraint given in eq. \eqref{eq:sys_mod}, such that their relative Euclidean distances are fixed.
Consequently, the covariance function is no longer a function of just two vectors, but also a function of the affine transform parameters including the rotation matrix angles, and the translation values.
How the transform parameters are to be incorporated appropriately in the elaborate design of the covariance function to maximize the differentiability between target nodes, is left to be investigated in the full article.

\subsection{\textbf{Towards Soft-connected Wireless Rigid Body Localization}}

Exemplified by the analysis of the \ac{SotA} \ac{RBL} techniques, a proposed technique coined \textit{''soft connected \acf{RBL}"} is a promising solution providing significant additional advantages in terms of precision, reliability, and applications.
In the following, we highlight the various open research possibilities for soft connected \ac{RBL} by describing how \ac{SotA} methods can be extended to enable the simultaneous localization of an entire sequence or arrange of rigid bodies.

So far, most of the \ac{SotA} \ac{RBL} works on just one object represented by one rigid body.
Therefore, in order to represent certain objects more realistically, we propose to represent certain objects by multiple rigid bodies with the main idea of soft-connecting these rigid bodies to achieve more diversity in the system and to create a soft-connection to group multiple rigid bodies to work on them simultaneously.

To that extent, consider the following examples.
Consider a vehicle such as a truck, which can be represented by two rigid bodies.
By viewing both of them simultaneously the localization can be improved, as well as further tracking and path prediction operations, as the connection of the two rigid bodies, provides a new set of constraints.
By leveraging soft connection algorithms that need to be developed, constraints such as distance between the rigid bodies, their corresponding angle, and more parameters can be optimized, or rather investigated in aforementioned applications to improve performance.
\ac{SotA} methods, such as \cite{PizzoICASSP2016} or \cite{Wu_22} already use information of other rigid bodies for their localization methods, which could further be extended by cooperative positioning techniques, considering the rigid bodies not as single independent targets but as one bigger object.

Another example considers ideas of localizing multiple rigid bodies simultaneously.
Examples of the former are, as mentioned before, platooning \cite{ZengTC2019} and swarming \cite{IbukiRAL2020} applications, as well as target tracking \cite{EckenhoffRAL2019}.
The areas of platooning and swarming can be extended by matrix completion approaches to create the soft connection of the multiple rigid bodies.
Focusing on tracking, time series methods can be applied to look at a tracking problem of one rigid body in many time slots, as one frame having multiple rigid bodies, where soft connections can be applied, improving tracking results.

We briefly provide a suggestion of the potential mathematical basis of enabling such soft connection between the \ac{SotA} rigid body framework.%
As the \ac{RBL} method imposes hard constraints between individual target nodes in the form of a fixed relative Euclidean distance, the proposed soft-connected rigid body aims to impose \ul{soft} constraints (\textit{i.e.,} inequality or bounded constraints instead of equality constraints) between each of the rigid body entities.
A soft connection between them can be included in the form of relative orientation between the rigid bodies, even in an anchorless scenario in \cite{PizzoICASSP2016} where multiple rigid bodies and their transformed position vectors are not given individually, but instead modeled jointly in a single transformation.
This model can be written as
\begin{equation}
\mathbf{S}\!=\![\mathbf{Q}_1|\mathbf{Q}_2]
\mleft[
\begin{array}{c|c}
\mathbf{C}_1 & \mathbf{0}_{2\times K} \\
\hline
\mathbf{0}_{2\times K} & \mathbf{C}_2
\end{array}
\mright]
+[\mathbf{t}_1|\mathbf{t}_2]
\mleft[
\begin{array}{c|c}
\mathbf{1}_{K\times 1} & \mathbf{0}_{K\times 1} \\
\hline
\mathbf{0}_{K\times 1} & \mathbf{1}_{K\times 1}
\end{array}
\mright],
\label{eq:sc}
\end{equation}
where $\mathbf{S}$ is the transformed position of both rigid bodies, $\mathbf{Q}_i$, $\mathbf{C}_i$, and $\mathbf{t}_i$ are the corresponding rotation, initial position, and translation vector for the $i$-th rigid body. 

Correspondingly, the joint estimation of $\mathbf{Q}_i$, $\mathbf{C}_i$, and $\mathbf{t}_i$ can be done with respect to a common constraint between each of the matrices.
Alternatively, the block structure of the matrices in eq. \eqref{eq:sc} can be modified such that the nullified blocks are replaced with covariance- or correlation-like matrices between each rigid body transformation, such that it can retain additional information on the \ul{relative} transformation between the bodies.


In light of the above conception of the potential mathematical modeling of the soft-connected rigid bodies, the full article will aim to deliver not only the thorough study of the \ac{SotA} methods leading towards the proposed idea, but an impactful formulation and contribution to enable the proposed method.

\end{document}